\documentclass[%
 reprint,
 amsmath,amssymb,
 aps,
prd
]{revtex4-2}

\usepackage{xcolor}
\usepackage{color,soul,enumitem}

\usepackage{graphicx}
\usepackage{dcolumn}
\usepackage{bm}

\newcommand{\dvol}{d\operatorname{vol}}
\newcommand{\tr}{\operatorname{tr}}

\begin{document}

\preprint{...} 

\title{MacDowell-Mansouri $\sigma$-model}

\author{Egor Pluzhnikov}
\email{ea.pluzhnikov@outlook.com}
\affiliation{ Faculty of Space Research, Lomonosov Moscow State University, Moscow, Russia }

\author{Stanislav Alexeyev}%
\email{alexeyevso@my.msu.ru}
\affiliation{
 Sternberg Astronomical Institute, Lomonosov Moscow State University, Moscow, Russia%
}
\affiliation{%
 Department of Quantum Theory and High Energy Physics, Physics Faculty,\\
 Lomonosov Moscow State University, Moscow, Russia}

\date{\today}

\begin{abstract}
Standard pre-geometric mechanisms reproduce the $\Lambda$CDM model but lack a natural source for inflation.  We propose a mechanism for the emergence of scalar-coupled topological terms by extending the Lie algebra of a MacDowell-Mansouri model to a Lie algebroid. These terms are commonly regarded as a source for cosmological inflation in effective string actions. With an additional constraint on torsion imposed, the theory reduces to a scalar-tensor theory belonging to the Horndeski class, and reproduces inflationary dynamics that meet existing constraints (Planck/BICEP/Keck and GW170817). 

\end{abstract}

\maketitle

\section{Introduction} According to the Big Bang models the energy scale drops from the Planck scale to string one and below, resulting first in compactification, moduli stabilization and supersymmetry breaking. In the infrared (IR) regime of string theory, various scalar fields coupled to topological terms appear \cite{Fernandes_2022}. These scalar fields drive the cosmological inflation, and the contribution of the topological terms is considered in effective, string-inspired actions modeling inflation \cite{Weinberg_2008, Kov_cs_2020, Odintsov_2018,odintsov2025}. Conversely, \textit{emergent/pre-geometric gravity} is expected to operate on the Planck scales \cite{Addazi_2025,addazi2025hamiltoniananalysispregeometricgravity}. A famous example is the MacDowell-Mansouri theory \cite{MM}. However, the existing emergent gravity framework derives only the $\Lambda$CDM model without the cosmological inflation source terms. The same effect occurs in the models with non-linear symmetry realization \cite{Alexeyev:2021lav,Alexeyev:2020lag}. In this article we report on the success in generating the scalar-tensor model containing the Gauss-Bonnet term with non-minimal coupling. The Lie algebra of the MacDowell-Mansouri theory is promoted to a Lie algebroid over the target space of a nonlinear $\sigma$-model \cite{Strobl_2004}, so its structure constant becomes a function of the scalar fields. The obtained model naturally features the source terms to drive inflation.

The paper is organized as follows. In Sec.~\ref{sec:review} we review the Lie algebroid Yang-Mills formalism, used intensively throughout the paper, the MacDowell-Mansouri theory, and its extensions based on the deformation of the underlying Lie algebra. The purpose of considering these extensions is to show how they make the topological terms manifest. Section~\ref{sec:NLSM_coupling} is devoted to the presentation of the model: its construction, its gauge symmetry and the active $SL(2,\mathbb{C})$ symmetry induced by vertical automorphisms of the Lie algebroid. In Sec.~\ref{sec:inflation} we discuss the application of the model to inflationary dynamics and how this symmetry relates equivalent descriptions. After the conclusions in Sec.~\ref{sec:conclusions}, there are three technical appendices. Appendix~\ref{app:expl_calc} is devoted to the calculation of the coupling functions between the scalars and the topological terms and to the self-dual representation of the model, analogous to the Gaillard-Zumino construction. Appendix~\ref{app:gauge} expands on the gauge invariance of the model, and in Appendix~\ref{app:superconstruction} we promote our model to a deformation of the original supersymmetric construction of MacDowell and Mansouri.

\section{Lie Algebroids and MacDowell-Mansouri theory}\label{sec:review}
To begin with, we outline the framework being used, namely, Lie algebroid Yang-Mills (LAYM) framework of Strobl \cite{Strobl_2004} and its extension to matter fields by Mayer and Strobl \cite{Strobl_2009}. A Lie algebroid $(E,\rho,[\cdot,\cdot])$ over a smooth manifold $\mathcal M$ (below, the target space of the $\sigma$-model) is a vector bundle $E\to\mathcal M$, a bundle map $\rho\colon E\to T\mathcal M$ (the anchor) and an $\mathbb R$-bilinear, antisymmetric bracket on $\Gamma(E)$ obeying the Jacobi identity for the Lie bracket and the Leibniz rule
\begin{equation}
  [s_1,fs_2]=f\,[s_1,s_2]+\bigl(\rho(s_1)f\bigr)\,s_2,\qquad f\in C^\infty(\mathcal M). \label{eq:leib}
\end{equation} 

In local coordinates $\phi^I$ on $\mathcal M$, $\partial_I=\partial/\partial\phi^I$, the anchor and the bracket are expanded in terms of a local frame $s_A \in \Gamma(E)$ of smooth sections of $E$ as $\rho(s_A)=\rho^I_A(\phi)\,\partial_I$ and
$[s_A,s_B]=C^C_{AB}(\phi)\,s_C$. We take $E=\mathcal M\times V$ trivial and $s_A$ a constant frame; in a general frame the formulas below acquire terms with the coefficients of a background connection on $E$ \cite{Strobl_2009}. Here, $C^C_{AB}=-C^C_{BA}$ and $\rho^I_A$ are referred to as structure functions. Expansion in this basis implies that the Jacobi identity holds if and only if
\begin{subequations}\label{eq:struct}
\begin{align}
  &\rho^J_A\partial_J\rho^I_B-\rho^J_B\partial_J\rho^I_A = C^C_{AB}\,\rho^I_C,\label{eq:str1}\\
  &C^E_{AD}C^D_{BC}+\rho^I_A\partial_IC^E_{BC}+\mathrm{cycl}(ABC)=0.\label{eq:str2}
\end{align}
\end{subequations}
Eq.~\eqref{eq:str1} states that the anchor is a homomorphism, $\rho([u,v])=[\rho(u),\rho(v)]$. In \eqref{eq:str2} the Jacobi identity acquires a derivative term along the anchor due to the $\phi$-dependence of structure functions. 

The main examples of algebroids are:
\begin{enumerate}[noitemsep]
    \item A Lie algebroid over a point $\mathcal{M} = \{\ \bullet\ \}$ is just a Lie algebra.
    \item Any Lie algebroid with a zero anchor $\rho = 0$ is a bundle of Lie algebras. In this case \eqref{eq:str1} holds trivially, \eqref{eq:str2} is the Jacobi identity at each point, and $C^{C}_{AB}(\phi)$ may depend on $\phi$ arbitrarily, provided the Jacobi identity holds pointwise.
    \item For $C^C_{AB}$ constant: $E=\mathcal
M\times\mathfrak g$ and, by \eqref{eq:str1}, $\rho\colon\mathfrak g\to \Gamma(T\mathcal M)$ is a Lie algebra homomorphism, i.e.\ an infinitesimal action. This construction is called an action Lie algebroid.
\end{enumerate}

The fields of the model are defined by a vector bundle morphism $a\colon TM\to E$ from the tangent bundle of spacetime $M$ to $E$. It covers a base map $\phi\colon M\to\mathcal M$ between the spacetime manifold $M$ and the target space $\mathcal{M}$ and is linear on the fibers, $a_x\colon T_xM\to E_{\phi(x)}$. The fiber part is a 1-form $A\in\Omega^1(M,\phi^*E)$, $a(v)=A^A(v)\,s_A(\phi(x))$ for $v\in T_xM$, so $a\equiv(\phi,A)$. Let $b^A$ be the frame of the dual bundle $E^*$, $b^A(s_B)=\delta^A_B$. The map $a$ pulls back functions on $\mathcal M$ and sections of $E^*$ and $\Lambda^2E^*$ to forms on $M$: $a^*f=f\circ\phi$, $(a^*\theta)(v)=\theta\bigl(a(v)\bigr)$ and $(a^*\omega)(v,w)=\omega\bigl(a(v),a(w)\bigr)$; in particular $a^*b^A=A^A$.

The algebroid differential ${}^Ed$ acts on functions
and on sections of $E^*$ by
\begin{align*}
  &({}^Edf)(u)=\rho(u)f,\\
  &({}^Ed\theta)(u,v)=\rho(u)\theta(v)-\rho(v)\theta(u)-\theta([u,v]),
\end{align*}
so that ${}^Ed\,b^A=-\tfrac12C^A_{BC}\,b^B\wedge b^C$. The field strength of $a$ is $\mathcal F=d\circ a^*-a^*\circ{}^Ed$. On the coordinate functions $\phi^I$ of $\mathcal M$, for which $a^*\phi^I=\phi^I(x)$ are the scalar fields, and on the dual frame
it gives
\begin{align*}
  &\mathcal F(\phi^I)=d\phi^I-\rho^I_A(\phi)A^A=\Delta^I,\\
  &\mathcal F(b^A)=dA^A+\tfrac12C^A_{BC}(\phi)A^B\wedge A^C=F^A.
\end{align*}
Geometrically, $\Delta=d\phi-\rho\circ a$ is the difference of two bundle maps $TM\to\phi^*T\mathcal M$: the differential of $\phi$ and the anchor composed with $a$. Note that $a$ is a Lie algebroid morphism iff $\Delta^I=0$ and $F^A=0$. Substituting $d\phi^J=\Delta^J+\rho^J_BA^B$ and using \eqref{eq:struct}, one finds the Bianchi identity
\begin{equation}
  d\Delta^I-\partial_J\rho^I_A\,A^A\wedge\Delta^J+\rho^I_AF^A=0 .
  \label{eq:bianchi}
\end{equation}
Therefore, the fields of the theory are identified with a map $\phi\colon M\to\mathcal M$ and a differential 1-form $A=A^A\,\phi^*s_A\in\Omega^1(M,\phi^*E)$.

The gauge transformations for the infinitesimal section $\varepsilon=\varepsilon^A\phi^*s_A\in\Gamma(\phi^*E)$ are 
\begin{equation}
\delta_\varepsilon\phi^I=\rho^I_A\varepsilon^A,\qquad
  \delta_\varepsilon A^A=d\varepsilon^A+C^A_{BC}(\phi)\,A^B\varepsilon^C.
  \label{eq:gt}
\end{equation}
Under it, the field strengths transform as
\begin{equation}
\begin{aligned}
\delta_\varepsilon\Delta^I&=\partial_J\rho^I_A\,\varepsilon^A\Delta^J,\\
  \delta_\varepsilon F^A&=C^A_{BC}F^B\varepsilon^C+\partial_IC^A_{BC}\,\Delta^I\wedge
  A^B\varepsilon^C,
\end{aligned}
\label{eq:dfs}
\end{equation}
where the second term of $\delta F^A$ is present only when the structure functions depend on
$\phi$.

If these both field strengths are squared, as in ordinary Yang-Mills theory, the derivative term $\partial_IC^A_{BC}\,\Delta^I\wedge A^B$ in $\delta_\varepsilon F^A$ cannot be compensated, and the preservation of the gauge invariance requires $\partial_I C^A_{BC} = 0$, i.e.\ ordinary Yang-Mills theory coupled to a $\sigma$-model \cite{Strobl_2009}. Instead, $\Delta^I=0$ is imposed by a Lagrange multiplier $B_I\in\Omega^{3}(M,\phi^*T^*\mathcal M)$, and only $F^A$ is squared with a fiber metric $g_{AB}(\phi)$:
\begin{equation}
  S_{\rm LAYM}=\int_MB_I\wedge\Delta^I-\frac12\int_Mg_{AB}(\phi)\,F^A\wedge*F^B .
  \label{eq:laym}
\end{equation}
The $B_I$ EOM is $\Delta^I=0$; by \eqref{eq:bianchi} it implies $\rho^I_AF^A=0$. All terms of the gauge variation proportional to $\Delta^I$ are compensated by a suitable transformation of $B_I$. The remaining part of the variation vanishes for a parameter along $s_C$ if
\begin{equation}
  \rho^I_C\partial_Ig_{AB}=C^D_{CA}g_{DB}+C^D_{CB}g_{AD},
  \label{eq:inv}
\end{equation}
which for $\rho^I_C\partial_Ig_{AB}=0$ is the ad-invariance of $g$ under $s_C$ \cite{Strobl_2009}. For the LAYM action see \cite{Strobl_2004,Strobl_2009}; for our model this is done in Appendix~\ref{app:gauge}.

Let us briefly recall the MacDowell-Mansouri construction. The reductive $SO(3,2)$-connection $A \in \Omega(M, \mathfrak{so}(3,2))$ is considered. It splits into the Lorentz and translation parts: $A = \frac12\omega^{ab} J_{ab} + e^a P_a$. The commutation relation $[P_a, P_b] = \lambda J_{ab}$ is modified. The splitting is transferred to a curvature:$F = \frac12\hat{F}^{ab} J_{ab} + T^a P_a$, where $\hat F^{ab} = R^{ab} + \lambda\ e^a \wedge e^b$ and $T$ is the torsion. The Yang-Mills type action $S = \int \mathrm{tr}( \hat F \wedge \star \hat F )$, where $\star$ is the inner Hodge star, results in the appearance of the cosmological constant term in the action \cite{MM,Wise_2010}. 

We use the following conventions. Lorentz indices are raised with the inner space metric $\eta_{ab}=\mathrm{diag}(-1,1,1,1)$. For the Levi-Civita symbol we fix $\epsilon_{0123}=-1$, hence $\epsilon^{0123}=+1$. The orientation is $\dvol=e^0\wedge e^1\wedge e^2\wedge e^3=\sqrt{-g}\,d^4x$, so that
$e^a\wedge e^b\wedge e^c\wedge e^d=\epsilon^{abcd}\,\dvol$. The spacetime Hodge star obeys
$\alpha\wedge*\beta=\langle\alpha,\beta\rangle\,\dvol$ on 1-forms, so
$d\varphi\wedge*d\varphi=g^{\mu\nu}\partial_\mu\varphi\partial_\nu\varphi\,\dvol$ and
$X=-\frac12(\partial\varphi)^2$ is the canonical kinetic term.

The Lie algebra is
\begin{equation}
\begin{aligned} {}
  [J_{ab},J_{cd}]&=\eta_{bc}J_{ad}-\eta_{ac}J_{bd}-\eta_{bd}J_{ac}+\eta_{ad}J_{bc},\\
  [J_{ab},P_c]&=\eta_{bc}P_a-\eta_{ac}P_b,\qquad [P_a,P_b]=h\,J_{ab}.
\end{aligned}
\end{equation}
with $h=\lambda$ for the MacDowell-Mansouri algebra: $h>0$ gives $\mathfrak{so}(3,2)$ (anti-de Sitter), $h<0$ gives $\mathfrak{so}(4,1)$ (de
Sitter), $h=0$ the Poincar\'e algebra. For complex $h\neq0$ the algebra is $\mathfrak{so}(5,\mathbb C)$, the common complexification of $\mathfrak{so}(3,2)$ and $\mathfrak{so}(4,1)$. Lie algebra valued forms are expanded in the basis, e.g. $X=\frac12X^{ab}J_{ab}$,
summed over all $a,b$. The connection $A=\frac12\omega^{ab}J_{ab}+e^aP_a$ has curvature
$F=dA+A\wedge A=\frac12\hat F^{ab}J_{ab}+T^aP_a$ with
\begin{equation}
\begin{aligned}
  &R^{ab}=d\omega^{ab}+\omega^a{}_c\wedge\omega^{cb},\qquad
   \hat F^{ab}=R^{ab}+h\,E^{ab},\\
  &T^a=De^a=de^a+\omega^a{}_b\wedge e^b,\qquad E^{ab}\equiv e^a\wedge e^b .
\end{aligned}
\end{equation}
In components $R^{ab}=\frac12R^{ab}{}_{cd}\,e^c\wedge e^d$, the Ricci tensor is $R^a{}_b=R^{ac}{}_{bc}$ and the scalar curvature is $R=R^{ab}{}_{ab}$. The first Bianchi identity reads $DT^a=R^a{}_b\wedge e^b$.
The Pontryagin, Holst and Gauss--Bonnet densities are
\begin{equation}
\begin{aligned}
  &R^{ab}\wedge R_{ab}=\mathcal P\,\dvol,\qquad R^{ab}\wedge e_a\wedge e_b=\mathcal H\,\dvol,\\
  &\mathcal G=R^2-4R_{ab}R^{ba}+R_{abcd}R^{cdab}.
\end{aligned}
\end{equation}
For the torsionless case $T=0$ the Bianchi identity gives
$R^{ab}\wedge e_b=0$, so $\mathcal H\equiv0$. 

The trace is taken in the vector representation of the Lorentz algebra
$\mathfrak{so}(3,1)$,
\begin{equation}
  \tr(J_{ab}J_{cd})=2(\eta_{ad}\eta_{bc}-\eta_{ac}\eta_{bd}),
\end{equation}
so that $\tr(X\wedge Y)=-X^{ab}\wedge Y_{ab}$.
The internal star is $\star J_{ab}=\frac12\epsilon_{abcd}J^{cd}$, $\star P_a=0$, so
$\star^2=-1$ on the Lorentz block and
\begin{equation}
  \tr(X\wedge\star Y)=-\tfrac12\epsilon_{abcd}X^{ab}\wedge Y^{cd}.
\end{equation}
On the Lorentz block $\tr(\star X\wedge\star Y)=-\tr(X\wedge Y)$ and $\tr(\star X\wedge
Y)=\tr(X\wedge\star Y)$ (note that both identities fail on the translations). Using these identities, the computation for the six traces yields
\begin{equation}
\begin{aligned}
  &\tr(R\wedge\star R)=\tfrac12\mathcal G\,\dvol, &&\tr(R\wedge R)=-\mathcal P\,\dvol,\\
  &\tr(E\wedge\star R)=R\,\dvol, &&\tr(R\wedge E)=-\mathcal H\,\dvol,\\
  &\tr(E\wedge\star E)=12\,\dvol, &&\tr(E\wedge E)=0.
\end{aligned}
\label{eq:dict}
\end{equation} 

A common approach is changing the Lie algebra of the MacDowell-Mansouri theory. The first example is subjecting it to a Wigner-In\"on\"u contraction \cite{Chirco_2025}, where in a limit $\lambda\to 0$ the algebra is reduced to an $\mathfrak{iso}(3,1)$. The second one is extending it to a complex parameter $\lambda \in \mathbb{C}$ \cite{Lazzarini}. As $\lambda$ would become a moduli-dependent function, and the physically interesting moduli are complex (e.g. the ones governing the deformations of $CY_3$), we consider $\lambda \in \mathbb{C}$. Making $\lambda$ complex, the commutation relations hold for the complexified Lie algebra $\mathfrak{g}_\mathbb{C} = \mathfrak{so}(3,2)\otimes \mathbb{C} = \mathfrak{so}(5,\mathbb{C})$. Therefore, $F \in \Omega^{2}(M,\mathfrak{g}_\mathbb{C})$. A natural candidate for the action is $S = \int [ \mathrm{tr}( \hat F \wedge \star \hat F) + \theta\mathrm{tr}(\hat F \wedge \hat F) + c.c.]$, where the $\theta$ term reveals the structure of topological terms. Expanding and using the traces \eqref{eq:dict}, it becomes
\begin{equation*}
    S = 2\int \sqrt{-g}[\tfrac12\mathcal{G} - \theta \mathcal{P} + 2\alpha R + 12(\alpha^2 - \beta^2) - 2\theta\alpha\mathcal{H}].
\end{equation*}
with $\lambda = \alpha + i\beta = re^{i\vartheta}$. The sign of the cosmological constant becomes governed by the phase $\vartheta$ and its value now depends upon the magnitude $r$. This provides a smooth interpolation between the dS, AdS, and Minkowski regimes \cite{moffat2026complexifiedspacetimemanifoldmapping}. 

The action could be extended by introducing fields valued in a direct sum $\mathfrak{g}=\bigoplus_{(i)=1}^m \mathfrak{g}_{\lambda_{(i)}}$ of copies of $\mathfrak{g}_{\lambda_{(i)}}$ with different $\lambda_{(i)}$ parameters (analogously to string axiverse \cite{Arvanitaki_2010} or the kinetic mixing of type IIB supergravity \cite{hebecker2024smallkineticmixingstring}).
\begin{equation*}
    S = \int M_{ij}\, \mathrm{tr}(\hat{F}^{(i)} \wedge \star \hat{F}^{(j)} ) + N_{ij} \mathrm{tr}(\hat F^{(i)} \wedge\hat F^{(j)}) + \mathrm{c.c.}
\end{equation*}
This action is equivalent to $\Lambda$CDM model, since the mixing matrices $M$ and $N$ (generalizing $\kappa_{MM}$ and $\theta$ parameters) are constant.  

\section{The Model}\label{sec:NLSM_coupling}
The Lie algebra of the MacDowell-Mansouri model is promoted to a Lie algebroid by promoting the parameter $\lambda$ to a field-dependent function $h$ on the moduli space \cite{Strobl_2004}. The \textit{moduli space} here is the target manifold of an arbitrary nonlinear $\sigma$-model, not fixed to a particular string compactification. 

Let $\mathcal{M}$ be a smooth manifold (it will become the target space of a nonlinear $\sigma$-model) and let $h : \mathcal{M} \to \mathbb{R}$ be a function.  We construct a Lie algebroid on the bundle $E = \mathcal{M} \times \mathfrak{g} \to \mathcal{M}$ by making the bracket of constant sections pointwise dependent $[P_a,P_b]_p = h(p)\,J_{ab}$, where $p\in\mathcal{M}$. The anchor $\rho : E \to T\mathcal{M}$ is defined on the standard constant sections by $\rho(s_{ab}) = \mathcal{J}_{ab}$ and $\rho(s_{a4}) = \mathcal{P}_a$, where $s_{ab} = \frac12 J_{ab}$, $s_{a4} = P_a$ and $\mathcal{J}_{ab},\mathcal{P}_a \in \Gamma^\infty(T\mathcal{M})$ are prescribed vector fields. The anchor is a homomorphism and the Jacobi identity holds iff $\mathcal{P}_a(h)=0$ and $\mathcal{J}_{ab}(h)=0$.

Let $\phi : M \to \mathcal{M}$ be a nonlinear $\sigma$-model, where $M$ is the spacetime manifold.  Pulling back the Lie algebroid $E$ along $\phi$ gives a Lie algebra bundle $\phi^*E \to M$, whose fibre at $x\in M$ is $E_{\phi(x)}\cong \mathfrak{g}$. The structure constants of Lie algebroid are the functions $C^a_{bc}(\phi(x))$ induced by the bracket on $E$. The only structure function depending on the point of $\mathcal{M}$ is $C^{cd}_{a4,b4}(\phi) = 2\,h(\phi)\,\delta^{cd}_{ab}$.

Writing $\rho(s_{AB}) = \rho^I_{AB}(\phi)\,\partial_I$ (indices $I,J,\dots$ on $\mathcal{M}$), the $1$-form $\Delta$ measuring the failure of $(\phi,A)$ to be a Lie algebroid morphism is $\Delta = d\phi - \rho(A) \in \Omega^1(M,\phi^*T\mathcal{M})$. In coordinate representation it takes the following form: $\Delta^I = d\phi^I - \rho^I_{AB}(\phi)\,A^{AB}$. Thus, the usual curvature $F = dA + \frac12[A,A]_\phi = \frac12 \hat F^{ab} s_{ab} + T^{a} s_{a4}$ is computed by fiber-wise bracket. Here,   $\hat F^{ab} = d\omega^{ab} + \omega^a{}_c\wedge\omega^{cb} + h(\phi)\,e^a\wedge e^b$ and $T^a = d_\omega e^a$ are exactly the components of the original MacDowell–Mansouri construction. Thus, the action is: 
\begin{eqnarray*}
   S = & + & \int_M U(\phi) B_I \wedge \Delta^I
      + \int M(\phi)\,\mathrm{tr}(\hat F\wedge \star \hat F) \\
      & + & \int N(\phi)\,\mathrm{tr}(\hat F\wedge \hat F) + \int g_{IJ}\Delta^I \wedge * \Delta^J,
\end{eqnarray*}
The B$\Delta$ term $B\wedge \Delta$ is introduced and the covariantized kinetic term $\Delta^I \wedge * \Delta^J$ with a usual spacetime Hodge star is used to make the action gauge-invariant \cite{Strobl_2004,Strobl_2009}. 

The potential $U(\phi)$ is introduced to provide the dynamics of the fields $\phi$. The reason is that $B$, being a Lagrange multiplier, constrains $U(\phi)\Delta^I = 0$. For $U\neq 0$, on the constraint surface the covariant $\sigma$-model kinetic term vanishes identically. The construction mirrors a standard Higgs term for $\sigma$-model \cite{percacci1998properties}. The metric $g_{IJ}$ on $\mathcal{M}$ is chosen to be invariant under the vector fields $\rho(s_{AB}) = \rho^I_{AB} (\phi)\partial_I$. The potential $U(\phi) = \frac12 g_{IJ}(\phi)\sum\nolimits_{A,B}\rho^I_{AB}(\phi)\rho^J_{AB}(\phi)$ generates the dynamics of scalars by relaxing the constraint $\Delta^I = 0$ and annihilates the anchor ($\rho = 0$ at the minimum). 

Consider the gauge transformations
\begin{equation*}
\begin{aligned}
    &\delta_\varepsilon\phi^I = \varepsilon^{AB}\rho^I_{AB}(\phi),\\
&\delta_\varepsilon A^{AB} = d\varepsilon^{AB} + C^{AB}_{CD,EF}(\phi)\,A^{CD}\varepsilon^{EF},
\end{aligned}
\end{equation*}
with the coupling matrices satisfying the anchor-invariance conditions
\begin{equation*}
\rho^I_{AB}\,\partial_I M_{ij} = 0,\qquad
\rho^I_{AB}\,\partial_I N_{ij} = 0,\qquad
\rho^I_{AB}\,\partial_I g_{JK}=0,
\end{equation*}
and the Lagrange multiplier $B_I$ assigned a suitable transformation law.
The gauge variation of the curvature generates terms proportional to the $\Delta^I$, because the structure constants depend on $\phi$.
These $\Delta$-dependent pieces are exactly compensated by an appropriate choice of $\delta_\varepsilon B_I$, namely
\[
\delta_\varepsilon B_I = -B_J\,\varepsilon^{AB}\partial_I\rho^J_{AB}
- \Psi^{\rm kin}_I - \Psi^{\rm top}_I,
\]
where $\Psi^{\rm kin}_I$ and $\Psi^{\rm top}_I$ are forms linear in $\Delta^I$ that arise from the gauge sectors. 
The covariantized $\sigma$-model kinetic term $\int g_{IJ}\,\Delta^I\wedge *\Delta^J$ can be added without breaking gauge invariance, provided the metric $g_{IJ}$ is anchor‑invariant and the transformation of $B_I$ is further modified by the corresponding term $\Psi^\sigma_I$.
Thus all $\Delta$-dependent terms cancel, and the full action is invariant off shell under local Lorentz transformations and diffeomorphisms. It is not invariant off shell under the translation part of the gauge algebra (nor, in Appendix~\ref{app:superconstruction}, under supersymmetry): a translation $\varepsilon^{a4}=\xi^a$ also rotates the Lorentz curvature into the torsion, $\delta_\xi\hat F^{ab}=h(\phi)\,(T^a\xi^b-T^b\xi^a)$, which contains no $\Delta^I$ and cannot be absorbed by $\delta_\varepsilon B_I$. These terms vanish on the torsion constraint $T=0$, where translations reduce to diffeomorphisms \cite{MM}. See the discussion in Appendix~\ref{app:gauge}.

\subsection{Final model}
This construction naturally commutes with the Lie algebra complexification $c_{\mathbb{C}}$ and the Whitney sum $\oplus$. Therefore, it is possible to introduce the connection taking values in a direct sum of $N$ copies of the complexified $\rho : E^{(i)}\to T\mathcal{M}$. The curvature 2-form is:
\begin{equation}
        F^{(i)} = \frac12 \bigl( R^{ab} + h^{(i)}(\phi, \bar\phi)\, e^a \wedge e^b \bigr) J_{ab} + T^a P_a.
\end{equation}
The natural action, extending the MacDowell-Mansouri construction by coupling it to a nonlinear $\sigma$-model is:
\begin{eqnarray}
        S & = & S_{B\Delta} + \int \mathrm{tr}\bigg[ M_{ij}(\phi)\, \hat{F}^{(i)} \wedge \star \overline{\hat{F}}{}^{(j)} \nonumber \\ 
        && + N_{ij}(\phi) \hat F^{(i)} \wedge\hat F^{(j)}  + \mathrm{c.c.} \bigg] \nonumber \\ && -  \frac{1}{2} \int g_{I\bar J} \Delta^I \wedge *  \,  \Delta^{\bar J}, 
\end{eqnarray}
where $X= -\frac{1}{2} \int g_{I\bar J} \Delta^I \wedge *  \,  \Delta^{\bar J}$ is a covariantized non-linear $\sigma$-model kinetic term, and the complex potential is considered: $S_{B\Delta}=\int [U(\phi,\bar\phi)B_I \wedge \Delta^I + \bar{U}(\phi, \bar\phi) \bar B_{\bar I} \wedge \Delta^{\bar I}] $. The matrices $M$ and $N$ are Hermitian, and complex symmetric correspondingly. The construction mirrors the Gaillard-Zumino construction and in case of real $h^{(i)}$, $M$ and $N$ e.g. for the supersymmetric completion of Appendix~\ref{app:superconstruction} can be as well rewritten in terms of self-dual curvature $\hat F^{(i)}_+ = \frac12( \hat F - \star i\hat{F})$. For a complex matrix $\frac14\mathcal{N}_{ij}(\phi) = N_{ij} + iM_{ij}$, the gauge part of the action is given by $\int \mathrm{Re}[ \mathcal{N}_{ij}\, \mathrm{tr}(\hat F^{(i)}_+ \wedge\hat F^{(j)}_+ )]$ (see Appendix \ref{app:expl_calc}).

\subsection{Two phases of the action}\label{sec:twophases}
In the unbroken phase the potential $U(\phi)\neq0$, so the field $B_I$ acts as a genuine Lagrange multiplier. Varying $B_I$ yields $\Delta^I = 0$. Subsequently, both the $B\Delta$ and kinetic terms vanish, and only the gauge part ($\int M F \wedge \star \bar F + N F \wedge F + c.c.$) of the action remains, supplemented by $\Delta^I \equiv d\phi^I - \rho^I_{AB}(\phi)A^{AB} = 0$ constraints. Taking the exterior derivative then yields $\rho^I_{AB}(\phi) F^{AB} = 0$. 

We split the index $I$ into $(a,\alpha)$ with $a=0,\dots,3$ and $\alpha=4,\dots,\dim\mathcal{M}-1$.  The anchor on the translation generators is assumed to be invertible in the $a$‑directions; in the adapted basis $\rho^a_{c4}=\delta^a_c$ (can always be done locally). Then, $e^a \equiv \Phi^a = (\rho^{-1})^a_I\bigl(d\phi^I - \rho^I_{cd}\,\omega^{cd}\bigr)$. Thus, the action (up to normalization coefficients) becomes:
\begin{equation}
\begin{aligned}
    S =   \int &\alpha(\phi^I)\epsilon_{abcd} \Phi^a \Phi^b \Phi^c \Phi^d +\beta(\phi^I)\mathcal{G} \nonumber \\
     &+ \gamma(\phi^I) \epsilon_{abcd} R^{ab} \Phi^c \Phi^d + \delta(\phi^I)R^{ab}\Phi_a\Phi_b \nonumber \\
     &+ \nu(\phi^I)R^{ab}R_{ab}.
\end{aligned}
\end{equation}
Collecting all contributions, the action in the unbroken phase reduces to a functional of the scalar fields $\phi^I$, the spin connection $\omega^{ab}$, and the curvature $R^{ab}$, with the tetrad eliminated. All coefficients $\alpha,\beta,\gamma,\delta,\nu$ and the functions $\rho^I_{cd}$ depend on the full set of moduli $\phi^I$, i.e.\ both $\phi^a$ and $\phi^\alpha$. Eliminating $d\phi^\alpha$ algebraically via the constraint $\Delta^I=0$ reduces the action to a theory of four scalars $\phi^{a}$ and the spin connection $\omega^{ab}$.

At the minimum of the potential, the $B\Delta$ term vanishes, and the field $B$ decouples, being unconstrained. The potential, due to the positive-definiteness of $g_{IJ}$, is minimized at $\rho = 0$. The gauge transformation of the scalar field become $\delta_\epsilon \phi = 0$, and the action remains invariant under the local Lorentz transformations of $A$ (Sec.~\ref{sec:NLSM_coupling}). All extra contributions from gauge transformations, previously canceled out by the $B\Delta$ term, become zero due to the dependence on $\delta_\epsilon \phi$.

This mechanism makes the scalar fields of the sigma model dynamical (leaving the theory invariant under local Lorentz transformations), since $\Delta \neq 0$ and the kinetic term of a $\sigma$-model also becomes nonzero. The action in this phase becomes 
\begin{eqnarray*}
  S & = & S_{\mathrm{kin}} +
  \int_{M_4}\mathrm{tr}\Bigl[
    \alpha(\phi)\,(e\wedge e)\wedge\star (e \wedge e)
    + \beta(\phi)\,R\wedge\star R \\
    & + & \gamma(\phi)\,R\wedge\star (e \wedge e)
    + \delta(\phi)\,R\wedge (e \wedge e)
    + \nu(\phi)\,R\wedge R \notag
  \Bigr] 
\end{eqnarray*}
with the coefficient functions given by (see Appendix \ref{app:expl_calc}):
\begin{equation*}
\begin{aligned}
  \alpha(\phi)  &= 2\sum_{i,j}\operatorname{Re}\bigl( M_{i\bar\jmath}\,h^{(i)}\bar h^{(\bar\jmath)} \bigr), &
  \beta(\phi)   &= 2\sum_{i,j}\operatorname{Re}M_{i\bar\jmath}, \\
  \gamma(\phi)  &= 2\sum_{i,j}\operatorname{Re}\bigl( M_{i\bar\jmath}\,(h^{(i)}+\bar h^{(\bar\jmath)}) \bigr), \\
  \delta(\phi)  &= 2\sum_{i,j}\operatorname{Re}\bigl( N_{ij}\,(h^{(i)}+h^{(j)}) \bigr), &
  \nu(\phi)     &= 2\sum_{i,j}\operatorname{Re}N_{ij}.
\end{aligned}
\label{eq:coeff}
\end{equation*}
By the traces \eqref{eq:dict}, this is
\begin{multline*}
  S = S_{\mathrm{kin}} + \int_{M_4}\sqrt{-g}\,\bigl[12\alpha(\phi) + \tfrac12\beta(\phi)\,\mathcal G
      + \gamma(\phi)\,R -\\
      -\delta(\phi)\,\mathcal H - \nu(\phi)\,\mathcal P\bigr]\,d^4x .
\end{multline*}. At the minimum of the $\sigma$-model Higgs potential $U(\phi)$ the Lie algebroid degenerates into the Lie algebra bundle with $\rho = 0$. The kinetic term became a standard sigma-model one: $X =-\frac12 g_{I\bar J} d\phi^I \wedge * d\phi^{\bar J}$.

\subsection{Global symmetry of the action}
The physical observables $\alpha,\beta,\dots$ (and thus, the action itself) are invariant under the transformation:
\begin{equation*}
h^{(i)} \to \frac{a h^{(i)}+b}{c h^{(i)}+d},\quad
M \to \overline{D} S^\dagger M S D,\quad
N \to D S^T N S D,
\end{equation*}
where $D = \operatorname{diag}(c h^{(i)}+d)$ and $S$ is a non‑degenerate matrix that acts on $\operatorname{span}\{\mathbf{h},\mathbf{1}\}$ as $\Lambda^{-1}=\left(\begin{smallmatrix} d & -b \\ -c & a \end{smallmatrix}\right)$ (and as the identity on the orthogonal complement). 

To understand where this symmetry originates, we recall that the bracket on the Whitney sum of Lie algebroids was defined as $[X^{(i)}, Y^{(j)}] = \delta^{ij}[X^{(i)}, Y^{(j)}]$ for any pair of the generators. Non-diagonal brackets can be defined as well: 
\begin{equation*}
\begin{aligned} {}
    &[P_a'^{(i)}, P_b'^{(j)}]' = \sum\nolimits_k H'^{ij}_k \; J_{ab}'^{(k)},\\
    &[J_{ab}'^{(i)}, P_c'^{(j)}]' = \sum\nolimits_k G'^{ij}_k \; \bigl( \eta_{bc} P_a'^{(k)} - \eta_{ac} P_b'^{(k)} \bigr)
\end{aligned}
\end{equation*}
Consider a vertical isomorphism $\Phi : E \to E'$ given by $\Phi(J_{ab}^{(i)}) = \sum_j U_{ij}\,J_{ab}'^{(j)}$ and $\Phi(P_a^{(i)}) = \sum_j V_{ij}\,P_a'^{(j)}$. Anchor preservation condition $\rho\circ \Phi = \rho$ yields $\sum\nolimits_{j} V_{ij} = 1$ and $\sum\nolimits_{j}U_{ij}=1$ for any $i$. Bracket preservation conditions $\Phi([X,Y]) = [\Phi(X), \Phi(Y)]'$ give:
\begin{equation*}
\begin{aligned}
&\sum\nolimits_{l,m} U_{il} V_{jm}\, G'^{lm}_k = \delta^{ij}\, V_{ik} \qquad \forall i,j,k. \\
&\sum\nolimits_{l,m} V_{il} V_{jm}\, H'^{lm}_k = \delta^{ij} h^{(i)} U_{ik} \qquad \forall i,j,k.
\end{aligned}
\end{equation*}
Thus, $\Phi$ is a Lie algebroid isomorphism if the conditions above hold. The connections transform as $A' = \sum\nolimits_k (u_k\omega^{ab} J'^{(k)}_{ab} + v_k e^a P'^{(k)}_a)$ for $u_k \equiv \sum_{i} U_{ik}$ and $v_k \equiv \sum_{i} V_{ik}$. For $u_k=1$, the Lorentz part of a curvature is $\hat F'^{(k)} = \frac12 \Big[ R^{ab} + \sum\nolimits_{i,j} v_i v_j H'^{ij}_k\, e^a\wedge e^b \Big] J'^{(k)}_{ab}$. For the diagonal case this means $\hat F_{\mathrm{diag}}^{(k)} = \frac12\bigl(R^{ab} + h'^{(k)} e^a\wedge e^b\bigr)J'^{(k)}_{ab}$, for the non-diagonal $\hat F'^{(k)} = \frac12\Big[ R^{ab} + W_k e^a \wedge e^b \Big] J'^{(k)}_{ab}$, where $W_k \equiv \sum_{i,j} v_i v_j H'^{ij}_k$. Equating $M_{kl}\, \hat F'^{(k)} \wedge \star \overline{\hat F}{}'^{(l)}$ and $M'_{kl}\, \hat F^{(k)}_{\mathrm{diag}} \wedge \star\overline{\hat F}{}^{(l)}_{\mathrm{diag}}$, one obtains the condition on bilinear forms $B_W \equiv (W,\mathbf{1})^\dagger M (W,\mathbf{1}) = (h',\mathbf{1})^\dagger M'(h', \mathbf{1}) \equiv B_{h'}$ for Hermitian $M$ and $M'$. Analogous computation gives the preservation of $A = (W,\mathbf{1})^T N (W,\mathbf{1})$. For $U = I$ and $V = (SD)^{-1}$ this gives $W_k = h^{(k)}$ (from the matrix form of $PP$ bracket condition) and $h'^{(i)} = \frac{ah^{(i)}+b}{ch^{(i)}+d}$, where $\Lambda = \left(\begin{smallmatrix}
    a & b \\ c & d
\end{smallmatrix}\right) \in SL(2,\mathbb{C})$. Therefore, the action possesses an \textit{active} $SL(2,\mathbb{C})$ symmetry, induced by vertical isomorphism of Lie algebroids with $U = I$ and $V = (SD)^{-1}$. This symmetry corresponds to different realizations of a Lie algebroid, similarly to the $SL(2,\mathbb{R})$-symmetry in type IIB supergravity formulation \cite{Bergshoeff_1996,Fernandez-Melgarejo2024ffg}. 

We nowhere assume that the matrices $U,V$ are constant on $\mathcal{M}$. In fact, the automorphism $\Phi$ may depend on the scalar fields $\phi$, provided that the Lie algebroid conditions hold pointwise. The only place where a non‑constant matrices could introduce additional terms is the computation of the Lorentz part of the curvature $\hat F'$. Under a field‑dependent $\Phi$, it transforms as $\hat F' = \Phi(\hat F) + \sum_k du_k \wedge \omega^{ab}J_{ab}^{(k)}$. The additional part vanishes for $u_k = 1$. Thus $\hat F' = \Phi(\hat F)$ exactly, with no extra pieces, even when $U$ and $V$ depend on $\phi$. Hence the full action is invariant under field‑dependent mappings.

\section{Inflationary Dynamics}\label{sec:inflation}
The model presented here can be promoted to a deformation of MacDowell and Mansouri's $D=4$, $\mathcal{N}=1$ anti-de Sitter supergravity coupled to moduli fields, by introducing a gravitino generator $Q_\alpha$ (see Appendix~\ref{app:superconstruction}); this requires real $h^{(i)}=4(m^{(i)})^2\geq0$. At a common mass $m^{(i)}=m$ and constant couplings it reproduces their theory and, with the constraint below, it is supersymmetric for any constant $N_{ij}$. We impose MacDowell and Mansouri's torsion constraint $R^{*i}=0$ \cite{MM}, the vanishing of the supertorsion, whose bosonic part is $T=0$. Thus, the action can be written in metric formalism
\begin{equation}
\begin{aligned}
S = \int_M \sqrt{-g}\Bigl[  &12\alpha(\phi) + X +\tfrac12\beta(\phi)\mathcal{G} + \gamma(\phi)R \\ 
&-  \delta(\phi)\mathcal{H} - \nu(\phi)\mathcal{P} 
\Bigr] \label{eq:Sfinal}
\end{aligned}
\end{equation}
with $R$ as scalar curvature, and $\mathcal{G}$, $\mathcal{H}$ and $\mathcal{P}$ are correspondingly Euler (Gauss-Bonnet), Holst and Pontryagin topological densities, nonminimally coupled to a $\sigma$-model. The action consists of five terms. The first two are exactly the potential and the kinetic terms of a $\sigma$-model. The $\beta$, $\delta$ and $\nu$ ones represent the contribution of the scalar-coupled topological terms. The term $\gamma(\phi)$ is the dilatonic one, making the gravitational constant dependent upon scalar fields. On the torsion constraint the Holst term is identically equal to zero due to the first Bianchi identity for torsionless connection: $e^a \wedge R_{ab} = 0$. The Pontryagin term, which for a running coefficient $\nu(\phi)$ gives dynamical Chern-Simons modified gravity \cite{Yunes_2009}, does not contribute to the cosmological background, since it vanishes identically for any spherically-symmetric metric, including FLRW \cite{Alexander_2015}. However, it produces a significant contribution to the primordial tensor perturbations, producing gravitational wave birefringence. This effect will be testable with future gravitational-wave detectors. 

The resulting action for the cosmological applications becomes
\begin{equation*}
S = \int \sqrt{-g}\,\bigl[12\alpha(\phi) + X + \tfrac12\beta(\phi) \mathcal{G} +  \gamma(\phi) R\bigr],
\end{equation*} representing a general scalar-tensor theory with non-minimal coupling containing source terms for inflation. 

Naturally, we identify the fields of the $\sigma$-model with the scalar fields of string theory, called moduli. In string picture, compactifications and symmetry reduction produce moduli fields, generating inflation. The landscape of compactification scenarios is extensive and the choice of the compactification scheme goes beyond the scope of this article. For the demonstration purposes we choose the simplest scenario: the inflationary dynamics happens on 1D background trajectory $\gamma \subset \mathcal{M}$, and the moduli become the functions of a natural parameter $\varphi$: $d\varphi^2 = g_{i\bar\jmath}\, d\phi^{i} d\phi^{\bar\jmath}$, which we effectively identify with an inflaton. Therefore, $d\phi^i = \frac{d\phi^i}{d\varphi}d\varphi$ and $d\bar\phi^{\bar\jmath} = \frac{d\bar\phi^{\bar\jmath}}{d\varphi}d\varphi$, and the kinetic part of the action becomes a canonical kinetic term $X = -\frac12 d\varphi \wedge \star d\varphi$ for the field $\varphi$. Such theories often suffer from Ostrogradsky instabilities. Therefore we verify that the model falls into the Horndeski class to avoid ghosts \cite{Kobayashi_2019}. For this, the equations of motion must be analyzed. The equations of motion can be obtained from the field equations of the theory of type $F(R, \varphi, X, \mathcal{G}, T)$, discarding $T$ dependence. The field equation for the metric is \cite{Kaczmarek:2020awp}:
\begin{equation*}
    \begin{aligned}\label{eq101} 
    G_{\mu\nu} F_R =& \frac{1}{2} g_{\mu\nu} (F - R F_R) +(g_{\mu\nu} \Box - \nabla_\mu\nabla_\nu) F_R + \\
    &+(\Xi_{\mu\nu} + \Delta_{\mu\nu}) F_{\mathcal{G}} + \nabla_\mu\varphi \nabla_\nu\varphi F_X.
\end{aligned}
\end{equation*}
Here, $F = 12\alpha + X +\tfrac12\beta \mathcal{G} + \gamma R$. The objects $\Xi_{\mu\nu}$ and $\Delta_{\mu\nu}$ are the corresponding tensor and differential operators resulting from the variation of Gauss-Bonnet term: 
\begin{align*}
&\begin{aligned}
\Xi_{\mu\nu} &= 2R R_{\mu\nu} - 4R^\alpha_\mu R_{\alpha\nu} - 4R_{\mu\alpha\nu\beta} R^{\alpha\beta} + 2R^\alpha_{\mu\beta\gamma} R_{\alpha\nu}^{\beta\gamma}
\end{aligned} \\
&\begin{aligned}
\Delta_{\mu\nu} &= 2R g_{\mu\nu} \square - 2R \nabla_\mu \nabla_\nu - 4g_{\mu\nu} R^{\alpha\beta} \nabla_\alpha \nabla_\beta - 4R_{\mu\nu} \square \\
&\quad + 4R^\alpha_\mu \nabla_\nu \nabla_\alpha + 4R^\alpha_\nu \nabla_\mu \nabla_\alpha + 4R_{\mu\alpha\nu\beta} \nabla^\alpha \nabla^\beta
\end{aligned}
\end{align*}
and the field equation of motion is just $\frac{1}{2} F_\varphi = \nabla_\mu (F_X \nabla^\mu \varphi) + F_X \Box\varphi$ \cite{Kaczmarek:2020awp}. Consider the higher-derivative terms $(g_{\mu\nu}\Box - \nabla_\mu\nabla_\nu)F_R$ and $\Delta_{\mu\nu} F_\mathcal{G}$ where $F_R = \gamma(\varphi)$ and $F_\mathcal{G} = \tfrac12\beta(\varphi)$. Since $\beta$ and $\gamma$ do not depend on the kinetic term, the Ostrogradsky instabilities (ghosts) are avoided and the parity-even part of the theory naturally lies in a subclass of Horndeski gravity \cite{Kobayashi_2019}. The parity-odd coupling $\nu(\phi)\mathcal{P}$ with a running coefficient is dynamical Chern-Simons gravity, which has third-order field equations and is not of Horndeski type; it does not act on the FLRW background and enters only the tensor perturbations. For constant $\nu$, as in the example below, the theory is of Horndeski type.

\subsection{An Example}
Choose a configuration with $12\alpha = -V(\varphi)$, $\beta = -\chi(\varphi)$, $\gamma = 1$. Then the action reproduces the well-studied scalar--Gauss--Bonnet model
\begin{equation}
S = \int d^4x\sqrt{-g}\left[ R + X - V(\varphi) - \tfrac12\chi(\varphi)\,\mathcal{G} \right].
\label{eq:SGB}
\end{equation}
This string-inspired model was previously studied by Odintsov and Oikonomou \cite{Odintsov_2018,odintsov2025}. It was shown that it supports a wide class of viable potentials and coupling functions. For example, one may consider the radion gauge potential $V(\varphi) = M(\kappa\varphi)^2/(d + k\varphi)$ and the coupling function $\chi(\varphi) = -\beta \varphi^{-3}$. In this case the slow-roll approximation gives $n_S \approx 0.974$, $r \approx 0.011$, and a gravitational wave speed $|c_T^2 - 1| < 10^{-15}$, in full agreement with Planck/BICEP/Keck data and the GW170817 constraint (see also other viable potentials in \cite{Odintsov_2018,odintsov2025}).

The knowledge of inflationary observables makes it possible to partially reconstruct the moduli-space parameters. For the simplest example above, one may take the number of copies as $N=2$ with $\rho = 0$. We trivially take $g_{IJ} = \delta_{IJ}$ on the moduli space $\mathcal{M} = \mathbb{R}^2$. The kinetic term becomes simply $S = -\frac{1}{2} \int d\phi^1 \wedge *d\phi^1 + d\phi^2 \wedge * d\phi^2$. We take $\gamma : \{ \phi^1 = \varphi, \phi^2 = 0 \}$ as the inflation trajectory, and thus the kinetic term becomes $X = -\frac12(\partial\varphi)^2$. In the basis $h^{(1)} = 0$, $h^{(2)} = 1$ (compatible with $h=4m^2\geq0$ of Appendix~\ref{app:superconstruction}), the matrix $B = (h,\mathbf{1})^T M (h,\mathbf{1})$, whose entries give $\alpha=2B_{11}$, $\beta=2B_{22}$ and $\gamma=4B_{12}$, is given by $B = \begin{pmatrix} -V(\varphi)/24 & 1/4 \\ 1/4 & - \chi(\varphi)/2 \end{pmatrix}$.
Thus, $M$ is
\begin{equation*}
\begin{aligned}
M &= ((h,\mathbf{1})^T)^{-1} B (h,\mathbf{1})^{-1} \\
  &= \frac{1}{24} \begin{pmatrix} -V - 12 - 12\chi & V + 6 \\ V + 6 & -V \end{pmatrix}
\end{aligned}
\end{equation*}
The matrix $N$ does not affect $\alpha,\beta,\gamma$, so it can be chosen, e.g., identically zero; this is the choice $\gamma_{\rm BI}=\infty$ of the Barbero-Immirzi parameter (Appendix~\ref{app:expl_calc}), and on the torsion constraint a constant $N$ is topological in any case.

Equivalent description can be given using the global automorphism symmetry. We apply a field-dependent transformation $\Lambda(\phi) \in SL(2,\mathbb{C})$ chosen such that the new matrix $M'$ has only one field-dependent entry: $M' = \frac14\left(\begin{smallmatrix} 0 & i \\ -i & -2\chi(\phi)\end{smallmatrix}\right) $. The transformed structure functions are determined by preservation of $B$, given by
\begin{equation*}
\begin{gathered}
i\bar h'^{(1)} - (i+2\chi)\bar h'^{(2)} = 1,\\
i\bar h'^{(1)}h'^{(2)} - i\bar h'^{(2)}h'^{(1)} - 2\chi|h'^{(2)}|^2 = -V/6 .
\end{gathered}
\end{equation*} The physical coefficients $\alpha,\beta,\gamma$ are unchanged, so the action remains the scalar--Gauss--Bonnet model. It follows that the two descriptions are related by a symmetry of the bosonic action and are therefore physically indistinguishable already at the fundamental level. 

The geometric origin of the active $SL(2,\mathbb{C})$-symmetry, combined with the fact that the construction admits a supersymmetric completion as a deformation of MacDowell–Mansouri supergravity, suggests that it may reflect a genuine redundancy of the hierarchically superior theory.
Thus we are led to the following \textit{hypothesis}:

\textit{In the landscape of string vacua, there exists a class of physically equivalent models connected by a symmetry of vertical automorphisms of a MMSM, which transforms the structure functions and pairings while leaving all inflationary observables invariant and uniting apparently distinct compactifications. }

\section{Conclusions} \label{sec:conclusions}
We suggest the extension of the MacDowell-Mansouri construction, coupled to a nonlinear $\sigma$-model. The underlying Lie algebra is promoted to a Lie algebroid. When the symmetry is broken by the potential term and the torsion constraint is added, it results in a scalar-tensor theory with a Gauss--Bonnet term with non-minimal coupling and other topological terms that naturally generates source terms for inflation. The resulting class of models satisfies current observational bounds and admits a direct geometric meaning. Moreover, it is the most general Gaillard-Zumino construction for the fields $\hat F$ and $\phi$ \cite{apruzzi2025gaillardzuminononinvertiblesymmetries}. Further development of this approach could allow deriving inflationary potentials from first principles, solving the inverse problem of calculating the compactification parameters from cosmological observables, and reproducing a Fab-Four type model, that naturally includes a bounce, genesis and inflation (see, for example, \cite{Zenin_2026}). 

\begin{acknowledgments}
The study was conducted under the state assignment of Lomonosov Moscow State University.
\end{acknowledgments}

\newpage
\appendix
\section{Gauge sector reduction}\label{app:expl_calc}

First, we calculate the dynamical term
\begin{multline}
  M_{i\bar\jmath}\,\hat{F}^{(i)}\wedge\star\overline{\hat{F}}{}^{(\jmath)} =
  M_{i\bar\jmath}\Bigl(R + h^{(i)}(e\wedge e)\Bigr)\wedge\star\Bigl(R + \bar h^{(\bar\jmath)}(e\wedge e)\Bigr)\\
  = M_{i\bar\jmath}\Bigl(
    R\wedge\star R
    + \bar h^{(\bar\jmath)}\,R\wedge\star (e\wedge e)\\
    + h^{(i)}\,(e\wedge e)\wedge\star R
    + h^{(i)}\bar h^{(\bar\jmath)}\,(e\wedge e)\wedge\star (e\wedge e)
  \Bigr).
  \label{eq:expM}
\end{multline}
Then using $R \wedge \star (e\wedge e) = (e\wedge e) \wedge \star R$ (Sec.~\ref{sec:review}) and adding the complex conjugate, for Hermitian $M$ we obtain
\begin{multline}
  \Bigl[M_{i\bar\jmath}\,\hat{F}^{(i)}\wedge\star\overline{\hat{F}}{}^{(\jmath)} + \mathrm{c.c.}\Bigr]
  = 2\bigl(\operatorname{Re}M_{i\bar\jmath}\bigr) R\wedge\star R \\
  + 2\operatorname{Re}\bigl( M_{i\bar\jmath}(h^{(i)}+\bar h^{(\bar\jmath)}) \bigr) R\wedge\star (e\wedge e) \\
  + 2\operatorname{Re}\bigl( M_{i\bar\jmath} h^{(i)}\bar h^{(\bar\jmath)} \bigr) (e\wedge e)\wedge\star (e\wedge e).
  \label{eq:Mterm}
\end{multline}
For $N\hat{F}\wedge \hat{F}$, the substitution of $\hat F$ yields the first part of the topological term:
\begin{equation}
  \begin{aligned}
  N_{ij}\,\hat{F}^{(i)}\wedge \hat{F}^{(j)} &=
  N_{ij}\Bigl(R + h^{(i)}(e\wedge e)\Bigr)\wedge\Bigl(R + h^{(j)}(e\wedge e)\Bigr)\\
  &= N_{ij}\Bigl(
    R\wedge R
    + (h^{(i)}+h^{(j)})\,R\wedge (e\wedge e) \\
    &\hphantom{=} + h^{(i)}h^{(j)}\,(e\wedge e)\wedge (e\wedge e)
  \Bigr),
  \end{aligned}
\end{equation}
thus
\begin{equation}
  \begin{aligned}
  \Big[N_{ij}\,\hat{F}^{(i)}\wedge \hat{F}^{(j)} + \mathrm{c.c.}\Big]
  &= 2\bigl(\operatorname{Re}N_{ij}\bigr) R\wedge R \\
  &\quad + 2\operatorname{Re}\bigl(N_{ij}(h^{(i)}+h^{(j)})\bigr) R\wedge (e\wedge e) \\
  &\quad + 2\operatorname{Re}\bigl(N_{ij}h^{(i)}h^{(j)}\bigr) (e\wedge e)\wedge (e\wedge e).
  \end{aligned}
  \label{eq:Nterm}
\end{equation}
The last term does not contribute, since $\tr[(e\wedge e)\wedge(e\wedge e)]=0$ by \eqref{eq:dict}. Summing over $i,j$, Eqs.~\eqref{eq:Mterm} and \eqref{eq:Nterm} give the coefficient functions $\alpha,\beta,\gamma,\delta,\nu$ of Sec.~\ref{sec:twophases}, and the traces \eqref{eq:dict} turn them into the coefficients $12\alpha$, $\frac12\beta$, $\gamma$, $-\delta$ and $-\nu$ of the densities $d\mathrm{vol}$, $\mathcal G$, $R$, $\mathcal H$ and $\mathcal P$.

For real structure functions $h^{(i)}$ and real matrices $M_{ij}$, $N_{ij}$ the
gauge part of the action of Sec.~\ref{sec:NLSM_coupling} can be rewritten in terms of the self-dual curvature
\begin{equation*}
  \hat F^{(i)}_+ = \frac12\bigl(\hat F^{(i)} - i\star\hat F^{(i)}\bigr),
\end{equation*}
where $\star$ is the inner Hodge star acting on the Lie algebra indices, $\star J_{ab}=\frac12\epsilon_{abcd}J^{cd}$ on the Lorentz part; on the translations $\star P_a=0$, and they do not enter the self-dual pairing.
We introduce a complex matrix $\mathcal{N}_{ij}$ related to $M$ and $N$ by $\mathcal{N}_{ij} = 4N_{ij} + 4i M_{ij}=X_{ij} + iY_{ij}$. Then the gauge sector is $S_{\text{gauge}} = \int \operatorname{Re}\bigl[
    \mathcal{N}_{ij}\, \operatorname{tr}\bigl(\hat F^{(i)}_+ \wedge \hat F^{(j)}_+\bigr) \bigr].$

To verify this, we first expand the trace of the product of two self-dual
curvatures. Using the definition and the properties $\star^2 = -1$,
$\operatorname{tr}(\star A \wedge \star B) = -\operatorname{tr}(A\wedge B)$
and $\operatorname{tr}(\star A \wedge B) = \operatorname{tr}(A \wedge \star B)$,
which hold on the Lorentz block (and on $\langle J\rangle\oplus\langle Q\rangle$ in Appendix~\ref{app:superconstruction}), we obtain
\begin{equation*}
  \begin{aligned}
  \operatorname{tr}\bigl(\hat F^{(i)}_+ \wedge \hat F^{(j)}_+\bigr)
  &= \frac14\operatorname{tr}\Bigl[
      \bigl(\hat F^{(i)} - i\star\hat F^{(i)}\bigr)
      \wedge \bigl(\hat F^{(j)} - i\star\hat F^{(j)}\bigr)
    \Bigr] \\
  &= \frac12\operatorname{tr}(\hat F^{(i)}\wedge\hat F^{(j)})
     - \frac{i}{2}\operatorname{tr}(\hat F^{(i)}\wedge\star\hat F^{(j)}).
  \end{aligned}
\end{equation*}
Now insert $\mathcal{N}_{ij} = X_{ij} + i Y_{ij}$:
\begin{equation*}
  \begin{aligned}
  &\mathcal{N}_{ij}\operatorname{tr}\bigl(\hat F^{(i)}_+ \wedge \hat F^{(j)}_+\bigr)
  = \tfrac12\Bigl[
      X_{ij}\operatorname{tr}(\hat F^{(i)}\wedge\hat F^{(j)})
      + Y_{ij}\operatorname{tr}(\hat F^{(i)}\wedge\star\hat F^{(j)})
    \Bigr] \\
  &\qquad + \tfrac{i}{2}\Bigl[
      Y_{ij}\operatorname{tr}(\hat F^{(i)}\wedge\hat F^{(j)})
      - X_{ij}\operatorname{tr}(\hat F^{(i)}\wedge\star\hat F^{(j)})
    \Bigr].
  \end{aligned}
\end{equation*}
Substituting $X_{ij}=4N_{ij}$ and $Y_{ij}=4M_{ij}$ into the real part we finally obtain
\begin{equation*}
  \begin{aligned}
  S_{\text{gauge}}
  &= \int \Bigl[
      2 N_{ij}\operatorname{tr}(\hat F^{(i)}\wedge\hat F^{(j)})
      + 2 M_{ij}\operatorname{tr}(\hat F^{(i)}\wedge\star\hat F^{(j)})
    \Bigr] \\
  &\equiv \int \operatorname{tr}\Bigl[
      M_{ij}\,\hat F^{(i)}\wedge\star\overline{\hat F}{}^{(j)}
      + N_{ij}\,\hat F^{(i)}\wedge\hat F^{(j)}
      + \mathrm{c.c.}
    \Bigr],
  \end{aligned}
\end{equation*}
where for real $\hat F^{(i)}$ we have
$\overline{\hat F}{}^{(i)} = \hat F^{(i)}$ and the complex conjugate
doubles the coefficients. The phase of $\mathcal{N}$ is the Barbero-Immirzi parameter: for one copy $\gamma_{\rm BI}=-M/N=-\tan\arg\mathcal{N}$, and the $\theta$ of Sec.~\ref{sec:review} equals $N/M=-1/\gamma_{\rm BI}$.

\section{Gauge invariance} \label{app:gauge}
The MacDowell–Mansouri action coupled to the $\sigma$-model is taken to be
\begin{equation}\label{eq:action}
  S = \int_M B_I \wedge \Delta^I
      + \int_M M(\phi)\,\mathrm{tr}(\hat F\wedge \star \hat F)
      + \int_M N(\phi)\,\mathrm{tr}(\hat F\wedge \hat F),
\end{equation}
where $B_I$ is a $(d-1)$-form Lagrange multiplier, $M,N$ are coupling ``matrices'' depending on $\phi$, and $\mathrm{tr}$ denotes the pairing of Appendix~\ref{app:expl_calc} (with the metric $g_{ab}$ in the $s_{AB}$ basis), taken on the Lorentz block; it is ad-invariant under $\mathfrak{so}(3,1)$ only, and so is its starred version, since $\star$ commutes with $\mathrm{ad}_J$ but not with $\mathrm{ad}_P$. It is necessary to add the $B\Delta$ term for the action to be gauge-invariant \cite{Strobl_2009}.

The $\Delta$-dependent part of the gauge variation of $S$ cancels provided the background fields satisfy
\begin{equation}\label{eq:invar}
  \rho^I_{AB}\,\partial_I M = 0,\qquad
  \rho^I_{AB}\,\partial_I N = 0,
\end{equation}
and the Lagrange multiplier transforms suitably. 

\paragraph*{Gauge transformations}
A gauge parameter is a section $\varepsilon \in \Gamma(\phi^*E)$, written $\varepsilon = \varepsilon^{AB} s_{AB}$ with $\varepsilon^{ab} = \omega_{\varepsilon}^{ab}$, $\varepsilon^{a4}= \xi^a$.  The fields transform as
\begin{equation}\label{eq:gtrans}
  \delta_\varepsilon \phi^I = \varepsilon^{AB}\rho^I_{AB}(\phi),\qquad
  \delta_\varepsilon A^{AB} = d\varepsilon^{AB} + C^{AB}_{CD,EF}(\phi)\,A^{CD}\varepsilon^{EF}.
\end{equation}

From these we compute the variation of $\Delta^I$:
\begin{align*}
  \delta_\varepsilon\Delta^I
  &= d(\delta_\varepsilon\phi^I) - \delta_\varepsilon(\rho^I_{AB}A^{AB})\nonumber\\
  &= d(\varepsilon^{AB}\rho^I_{AB}) 
     - \varepsilon^{CD}\rho^J_{CD}\,\partial_J\rho^I_{AB}\,A^{AB}
     - \rho^I_{AB}\,\delta_\varepsilon A^{AB}\nonumber\\
  &= \varepsilon^{AB}\partial_J\rho^I_{AB}\,d\phi^J\\
     &\qquad - \varepsilon^{AB}\rho^J_{AB}\,\partial_J\rho^I_{CD}\,A^{CD}\\
     &\qquad - \rho^I_{AB}\,C^{AB}_{CD,EF}\,A^{CD}\varepsilon^{EF},
\end{align*}
where in the last line the $d\varepsilon$ terms have cancelled.  Substituting $d\phi^J = \Delta^J + \rho^J_{KL}A^{KL}$ and using the structural identity
\[
  \rho^J_{AB}\,\partial_J\rho^I_{CD} - \rho^J_{CD}\,\partial_J\rho^I_{AB}
  = C^{EF}_{AB,CD}\,\rho^I_{EF},
\]
which follows from the anchor being a homomorphism, the $A^{CD}$‑dependent terms cancel exactly.  One obtains
\begin{equation}\label{eq:varDelta}
  \delta_\varepsilon\Delta^I = \varepsilon^{AB}\,\partial_J\rho^I_{AB}\,\Delta^J.
\end{equation}

The variation of the curvature $F^{AB}$ is
\begin{align}
  \delta_\varepsilon F^{AB}
  &= d(\delta_\varepsilon A^{AB})
     + \tfrac12\bigl(\delta_\varepsilon C^{AB}_{CD,EF}\bigr)A^{CD}\wedge A^{EF}\\
     &\qquad
     + C^{AB}_{CD,EF}\,\delta_\varepsilon A^{CD}\wedge A^{EF}\nonumber\\
  &= C^{AB}_{CD,EF}\,\varepsilon^{CD}F^{EF}
     + \partial_I C^{AB}_{CD,EF}\,\Delta^I\wedge A^{CD}\varepsilon^{EF}.
  \label{eq:varF}
\end{align}
The first term is the covariant variation $[\varepsilon,F]^{AB}$, while the second one is present only for those components whose structure constants depend on $\phi$.  In our basis the only $\phi$‑dependent constant is $C^{ab}_{c4,d4} = 2h(\phi)\delta^{ab}_{cd}$, therefore
\[
  \partial_I C^{ab}_{CD,EF}\,\Delta^I\wedge A^{CD}\varepsilon^{EF}
  = 2\,\partial_I h\,\Delta^I\wedge (e^a\xi^b - e^b\xi^a).
\]
The components $F^{a4}$ have no $\Delta$‑dependent variation.

\paragraph*{Action invariance}
Let $\delta_\varepsilon$ act on $S$.  The variation of the $B\Delta$ term is
\[
  \delta_\varepsilon\!\int_M B_I\wedge\Delta^I
  = \int_M \bigl(\delta_\varepsilon B_I + B_J\,\varepsilon^{AB}\partial_I\rho^J_{AB}\bigr)\wedge\Delta^I.
\]
For the kinetic term, using $\delta_\varepsilon M=0$ and leaving aside the covariant part $[\varepsilon,F]$ of $\delta_\varepsilon F$ (treated below), we obtain
\[
  \delta_\varepsilon S_{\rm kin}
  = \int_M M(\phi)\,\bigl[\mathrm{tr}(D\wedge\star \hat F) + \mathrm{tr}(\hat F\wedge\star D)\bigr],
\]
where $D^{AB} = \partial_I C^{AB}_{CD,EF}\,\Delta^I\wedge A^{CD}\varepsilon^{EF}$ is the $\Delta$‑dependent part of $\delta_\varepsilon F$ from \eqref{eq:varF}.  Writing the trace in components gives
\[
  \mathrm{tr}(D\wedge\star \hat F) = g_{ab}\,D^a\wedge\star \hat F^b,\qquad
  \mathrm{tr}(\hat F\wedge\star D) = g_{ab}\,\hat F^a\wedge\star D^b.
\]
Because $\star$ acts only on the algebraic indices, $\star(\Delta^I\wedge X) = \Delta^I\wedge\star X$, and we can factor out $\Delta^I$:
\[
  \delta_\varepsilon S_{\rm kin}\big|_\Delta
  = \int_M \Delta^I\wedge \Psi^{\rm kin}_I,
\]
with
\[
  \Psi^{\rm kin}_I = M\,g_{ab}\Bigl(
     \partial_I C^{a}_{cd}\,\varepsilon^d A^c\wedge\star \hat F^b
   + \partial_I C^{b}_{cd}\,\varepsilon^d\,\hat F^a\wedge\star A^c \Bigr).
\]
An identical computation for the topological term (using $\delta_\varepsilon N=0$) yields
\[
  \delta_\varepsilon S_{\rm top}\big|_\Delta
  = \int_M \Delta^I\wedge \Psi^{\rm top}_I,
\]
\[
  \Psi^{\rm top}_I = N\,g_{ab}\Bigl(
     \partial_I C^{a}_{cd}\,\varepsilon^d A^c\wedge \hat F^b
   + \partial_I C^{b}_{cd}\,\varepsilon^d\,\hat F^a\wedge A^c \Bigr).
\]

Collecting all $\Delta$‑dependent terms, the total variation of the action is
\[
  \delta_\varepsilon S = \int_M \Bigl(
      \delta_\varepsilon B_I
      + B_J\,\varepsilon^{AB}\partial_I\rho^J_{AB}
      + \Psi^{\rm kin}_I
      + \Psi^{\rm top}_I
    \Bigr)\wedge \Delta^I.
\]
Thus, if we \textit{define} the gauge transformation of the Lagrange multiplier to be
\begin{equation}\label{eq:deltaB}
  \delta_\varepsilon B_I :=
    -\,B_J\,\varepsilon^{AB}\partial_I\rho^J_{AB}
    - \Psi^{\rm kin}_I
    - \Psi^{\rm top}_I,
\end{equation}
all $\Delta$-dependent terms cancel for any $\varepsilon$. The covariant part cancels for Lorentz parameters, since the pairing is ad-invariant under $\mathfrak{so}(3,1)$ and $\star$ commutes with $\mathrm{ad}_J$. However, this does not hold for $\mathrm{ad}_P$. For a translation $\varepsilon=\xi^aP_a$, $\delta_\varepsilon A=d\varepsilon+[A,\varepsilon]$ gives $\delta e^a=D\xi^a$, $\delta\omega^{ab}=h(e^a\xi^b-e^b\xi^a)$ and, up to $\Delta$-dependent terms,
\[
  \delta_\xi\hat F^{ab}=h\,(T^a\xi^b-T^b\xi^a),
\]
so that $\delta_\xi(\epsilon_{abcd}\hat F^{ab}\wedge\hat F^{cd})=4h\,\epsilon_{abcd}T^a\xi^b\wedge\hat F^{cd}$ and $\delta_\xi(\hat F^{ab}\wedge\hat F_{ab})=4h\,T^a\xi^b\wedge\hat F_{ab}$. These terms are linear in the torsion and contain no $\Delta^I$, so $\delta_\varepsilon B_I$ cannot absorb them. The action is therefore invariant off shell under local Lorentz transformations and diffeomorphisms; translations, and (in Appendix~\ref{app:superconstruction}) supersymmetry, are broken off shell by terms linear in $T$. On the torsion constraint translations reduce to diffeomorphisms, and the supersymmetric completion is supersymmetric at a common mass of the copies.

\paragraph*{Remark on the $\sigma$-model kinetic term.}
The action \eqref{eq:action} does not contain a genuine kinetic term for the scalar fields $\phi^I$.  The usual $\sigma$-model term
\[
  S_{\sigma} = -\frac12\int_M g_{IJ}(\phi)\,d\phi^I\wedge *\,d\phi^J ,
\]
where $*$ is the spacetime Hodge star, is not gauge‑invariant.  Indeed, using
$\delta_\varepsilon\phi^I = \varepsilon^{AB}\rho^I_{AB}$,
\[
  \delta_\varepsilon S_{\sigma}
  = -\frac12\int (\delta_\varepsilon g_{IJ})\,d\phi^I\wedge *d\phi^J
    -\int g_{IJ}\, \delta_\varepsilon(d\phi^I)\wedge *d\phi^J .
\]
With $\delta_\varepsilon(d\phi^I)=d(\delta_\varepsilon\phi^I)= d\varepsilon^{AB}\,\rho^I_{AB}+\varepsilon^{AB}\partial_J\rho^I_{AB}\,d\phi^J$, the second integral produces a term linear in $d\varepsilon^{AB}$,
\[
  -\int g_{IJ}\,\rho^I_{AB}\,d\varepsilon^{AB}\wedge *d\phi^J ,
\]
which cannot be cancelled by any other term in the action.  Hence $S_{\sigma}$ is not invariant.

One may try to remedy this by replacing $d\phi^I$ with its covariant counterpart $\Delta^I$:
\[
  S_{\Delta} = -\frac12\int_M g_{IJ}(\phi)\,\Delta^I\wedge *\Delta^J .
\]
Because $\delta_\varepsilon\Delta^I = \varepsilon^{AB}\partial_J\rho^I_{AB}\,\Delta^J$ (eq.~\ref{eq:varDelta}), the variation of $S_{\Delta}$ is
\[
  \delta_\varepsilon S_{\Delta}
  = -\int g_{IJ}\,\varepsilon^{AB}\partial_K\rho^I_{AB}\,\Delta^K\wedge *\Delta^J
    -\frac12\int (\delta_\varepsilon g_{IJ})\,\Delta^I\wedge *\Delta^J .
\]
Under the invariance hypothesis $\rho^I_{AB}\partial_I g_{JK}=0$ (which is the same as $\delta_\varepsilon g_{IJ}=0$), the second term vanishes and the first one becomes
\[
  \delta_\varepsilon S_{\Delta}
  = \int \Delta^I \wedge \bigl( - g_{KJ}\,\varepsilon^{AB}\partial_I\rho^K_{AB} *\Delta^J \bigr)
  \equiv \int \Delta^I \wedge \Psi^{\sigma}_I .
\]
The form $\Psi^{\sigma}_I$ is linear in $\Delta$ and can, in principle, be absorbed into the definition \eqref{eq:deltaB} of $\delta_\varepsilon B_I$.  This would preserve the cancellation of the $\Delta$-dependent terms with $S_{\Delta}$ added.

However, the term $\int B_I\wedge\Delta^I$ in \eqref{eq:action} acts as a Lagrange multiplier.  Its equation of motion is
\[
  \frac{\delta S}{\delta B_I} = 0 \quad\Longrightarrow\quad \Delta^I = 0 .
\]
On this constraint surface the covariant $\sigma$-model term vanishes identically ($\Delta^I=0$), and the scalar fields become covariantly constant ($d\phi^I = \rho^I_{AB}A^{AB}$).

Therefore, a natural candidate for the action is
\begin{equation}
\begin{aligned}
  S =
      & \int_M M(\phi)\,\mathrm{tr}(\hat F\wedge \star\hat  F)+ \int_M N(\phi)\,\mathrm{tr}(\hat F\wedge \hat F) \\
      &\qquad +\int_M V(\phi)\, B_I \wedge \Delta^I -\frac12\int_M g_{IJ}(\phi) \Delta^I \wedge *\Delta^J,
\end{aligned}
\end{equation}
where $M(\phi)$, $N(\phi)$ and $g_{IJ}$ are anchor-invariant \eqref{eq:invar}, and $V(\phi)$ is a Higgs-like potential, that constraints $\Delta^I = 0$ in unbroken phase and generates the $\sigma$-model dynamics at the minimum $V(\phi) = 0$, where the gauge symmetry is broken.

\section{Supersymmetric Construction}
\label{app:superconstruction}

In this appendix we provide a self-contained, explicit derivation of the full supersymmetric action that couples $N=1$ supergravity (Rarita--Schwinger field) to a non-linear $\sigma$-model and topological terms. We promote
the structural constants of the $\mathfrak{osp}(1,4)$ superalgebra to field-dependent functions over a moduli space $\mathcal{M}$.

Let $\mathcal{M}$ be a complex manifold (the target space of the $\sigma$-model) with local coordinates $\phi^I$. Over $\mathcal{M}$ we consider a trivial vector bundle $E=\mathcal{M}\times\mathfrak{g}$, where $\mathfrak{g}=\mathfrak{osp}(1,4)$ is the graded Lie algebra with even generators $J_{ab}$ (Lorentz), $P_a$ (de Sitter translations) and odd generator $Q_\alpha$ (Majorana spinor). The anchor is a vector-bundle morphism $\rho:E\to T\mathcal{M}$. A Lie superalgebroid structure on $E$ is defined by a field-dependent bracket on the space of constant sections, which we write as ($p\in\mathcal{M}$):
\begin{align}{}
[J_{ab},J_{cd}]_p &= \eta_{bc}J_{ad}-\eta_{ac}J_{bd}-\eta_{bd}J_{ac}+\eta_{ad}J_{bc},\nonumber\\
[J_{ab},P_c]_p &= \eta_{bc}P_a-\eta_{ac}P_b,\quad
[J_{ab},Q_\alpha]_p = \tfrac12 Q_\beta(\gamma_{ab})^{\beta}{}_{\alpha},\nonumber\\
[P_a,P_b]_p &= h(p)\,J_{ab},\nonumber\\
[P_a,Q_\alpha]_p &= m(p)\,Q_\beta(\gamma_{a})^{\beta}{}_{\alpha},\\
\{Q_\alpha,Q_\beta\}_p &= -\tfrac12 k(p)\sum\nolimits_{a<b}(C\gamma^{ab})_{\alpha\beta}J_{ab}\nonumber\\
&\quad+\tfrac12 \ell(p)\,(C\gamma^{a})_{\alpha\beta}P_a,\nonumber
\end{align}
with $h,k,\ell,m$ real functions on $\mathcal{M}$, $C$ the charge conjugation matrix, and the
gamma matrices satisfy $\{\gamma_a,\gamma_b\}=2\eta_{ab}$, $\eta_{ab}=\operatorname{diag}(-1,1,1,1)$, in a Majorana representation, with $\bar\psi=\psi^TC$ and $\gamma_5=i\gamma^0\gamma^1\gamma^2\gamma^3$. From the $(J,J,Q)$ and $(J,P,Q)$ super-Jacobi identities, the consistent spinor representation $\sigma$ is $\sigma(J_{ab})=\tfrac12\gamma_{ab}$, $\sigma(P_a)=m\gamma_a$, with the index order $[X,Q_\alpha]=Q_\beta\,\sigma(X)^\beta{}_\alpha$. For simplicity, we consider the minimized potential phase, so $\rho \equiv 0$ from the beginning. The Jacobi identities for the bracket and the compatibility with the anchor are then satisfied trivially.
The super-Jacobi identities between the brackets restrict the functions: the $(P_a,P_b,Q_\alpha)$ yields $h = 4 m^2$, then $(P_a,Q_\alpha,Q_\beta)$ fixes the ratio $k/\ell = 2m$, so $\ell$ itself is a normalization of $Q$ (we set $k=2$, so that $\ell=1/m$). The superalgebroid is controlled by a single real function $m(\phi)$ on the moduli space. 

We introduce $N$ copies of the superalgebroid, labelled by $i,j=1,\dots,N$, with structural
functions $h^{(i)}=4(m^{(i)})^2$, $k^{(i)}=2$, $\ell^{(i)}=1/m^{(i)}$: a single invariant pairing for all copies (below) requires $k^{(i)}$ to be copy-independent, and hence $\ell^{(i)}\propto1/m^{(i)}$. The gauge connection
is a one-form on spacetime $M_4$ taking values in the direct sum
$\bigoplus_{i=1}^N \mathfrak{g}^{(i)}$: $A = \sum_i \left( \tfrac12 \omega^{ab} J_{ab}^{(i)} + e^a P_a^{(i)} + \psi^\alpha Q_\alpha^{(i)} \right),$ where $\omega^{ab}$ is the spin connection, $e^a$ the vierbein, and $\psi^\alpha$ the (single) Majorana Rarita--Schwinger field. The curvature is then $F^{(i)} = \tfrac12 \hat{F}^{(i)ab} J_{ab}^{(i)} + T^{(i)a} P_a^{(i)} + \rho^{(i)\alpha} Q_\alpha^{(i)},$ with
\begin{align}
\hat{F}^{(i)ab} &= R^{ab} + h^{(i)} e^a\!\wedge\! e^b - \tfrac12\bar\psi\!\wedge\!\gamma^{ab}\psi ,\nonumber\\
T^{(i)a} &= d_\omega e^a + \tfrac{1}{4m^{(i)}} \bar\psi\!\wedge\!\gamma^a\psi ,\nonumber\\
\rho^{(i)\alpha} &= D\psi^\alpha + m^{(i)} e^a\!\wedge\!(\gamma_a\psi)^\alpha ,\\
R^{ab} &= d\omega^{ab} + \omega^{a}{}_c\!\wedge\!\omega^{cb},\quad
D\psi^\alpha = d\psi^\alpha + \tfrac14 \omega^{ab}\!\wedge\!(\gamma_{ab}\psi)^\alpha .\nonumber
\end{align}
We extend the trace form to the odd generators:
\begin{equation}
\begin{aligned}
\operatorname{tr}(J_{ab}^{(i)}J_{cd}^{(j)}) &= 2(\eta_{ad}\eta_{bc}-\eta_{ac}\eta_{bd}),\\
\operatorname{tr}(Q_\alpha^{(i)}Q_\beta^{(j)}) &= -4\,C_{\alpha\beta},
\end{aligned}
\label{eq:str}
\end{equation}
$\operatorname{tr}(J,Q)=0$, the same for every pair of copies $i,j$ (a factor $\delta^{ij}$ would remove every off-diagonal $M_{ij}$, $N_{ij}$). 

Let $M_{ij}(\phi)$ and $N_{ij}(\phi)$ be real symmetric
matrices on $\mathcal{M}$ invariant under the vector field $\rho_{AB}^I\partial_I$. Introduce:
\begin{equation}
\mathcal{N}_{ij}(\phi) = 4\bigl(N_{ij}(\phi) + i\,M_{ij}(\phi)\bigr).
\label{eq:Nmatrix}
\end{equation}
Define the self-dual part of the curvature with respect to the Hodge star on the Lie algebra
indices: $\star$ acts as $\star J_{ab}= \tfrac12\varepsilon_{abcd}J^{cd}$,
$\star Q_\alpha = Q_\beta(-i\gamma_5)^\beta{}_\alpha$, $\star P_a=0$, and satisfies $\star^2 = -1$ on $\langle J\rangle\oplus\langle Q\rangle=\mathfrak{osp}(1,2\mathbb{C})$. Then
\begin{equation}
F_+^{(i)} = \tfrac12 (F^{(i)} - i\star F^{(i)}).
\label{eq:selfdual}
\end{equation}
It is projected on $\mathfrak{osp}(1,2\mathbb{C})$: the $P$ component, on which $\star$ vanishes, is dropped. The action we propose is
\begin{multline}
S = \int_{M_4} \operatorname{Re}\!\Big[ \mathcal{N}_{ij}(\phi)\;
\operatorname{tr}\!\big( F_+^{(i)} \wedge F_+^{(j)} \big) \Big] -\frac12 \int g_{I\bar J} d\phi^I \wedge * d\phi^J.
\label{eq:gaugeaction}
\end{multline}
Here, the matrix $\mathcal{N}_{ij}$ is identified with the moduli kinetic matrix of the string theory. Complex conjugation acts as $(\theta_1\theta_2)^*=\theta_1^*\theta_2^*$; equivalently, the gauge part is defined by its real form $\int[2N_{ij}\operatorname{tr}(F^{(i)}\wedge F^{(j)})+2M_{ij}\operatorname{tr}(F^{(i)}\wedge\star F^{(j)})]$ on $\mathfrak{osp}(1,2\mathbb{C})$. With the supertorsion constraint $T^{(i)a}=0$ imposed, the action at a common mass $m^{(i)}=m$ is supersymmetric for any constant $M_{ij}$, $N_{ij}$.
\begin{widetext}
Expanding the action yields all terms of the supersymmetric theory, both bosonic and fermionic, with coefficients determined by the field combinations
\[
\begin{gathered}
\Sigma_M=\sum_{ij}M_{ij},\quad \Sigma_N=\sum_{ij}N_{ij},\quad \Sigma_{Mm}=\sum_{ij}M_{ij}(m_i+m_j),\quad \Sigma_{Nm}=\sum_{ij}N_{ij}(m_i+m_j),\\
\Pi=\sum_{ij}M_{ij}m_i^2m_j^2,\quad \Sigma_{M(m^2)}=\sum_{ij}M_{ij}(m_i^2+m_j^2),\quad \Sigma_{N(m^2)}=\sum_{ij}N_{ij}(m_i^2+m_j^2),\\
\widetilde M=\sum_{ij}M_{ij}m_im_j,\quad \widetilde N=\sum_{ij}N_{ij}m_im_j,
\end{gathered}
\]
with also $S_M=\Sigma_{M(m^2)}+2\widetilde M$ and $S_N=\Sigma_{N(m^2)}-2\widetilde N$. The action is given by the Lagrangian $L=L^{(0)}+L^{(2)}+L^{(4)}$:
\begin{align*}
L^{(0)}={}&-\Sigma_M\,\epsilon_{abcd}\,R^{ab}\wedge R^{cd}
-4\Sigma_{M(m^2)}\,\epsilon_{abcd}\,R^{ab}\wedge e^c\wedge e^d
-16\,\Pi\,\epsilon_{abcd}\,e^a\wedge e^b\wedge e^c\wedge e^d\\
&-2\Sigma_N\,R^{ab}\wedge R_{ab}
-8\Sigma_{N(m^2)}\,R^{ab}\wedge e_a\wedge e_b ,\\[6pt]
L^{(2)}={}&\Sigma_M\,\epsilon_{abcd}\,R^{ab}\wedge\bar\psi\wedge\gamma^{cd}\psi
+2\Sigma_N\,R^{ab}\wedge\bar\psi\wedge\gamma_{ab}\psi\\
&+2\Sigma_{M(m^2)}\,\epsilon_{abcd}\,e^a\wedge e^b\wedge\bar\psi\wedge\gamma^{cd}\psi
+4\Sigma_{N(m^2)}\,e^a\wedge e^b\wedge\bar\psi\wedge\gamma_{ab}\psi\\
&+8i\,\Sigma_M\,\overline{D\psi}\wedge\gamma_5\,D\psi
-8\,\Sigma_N\,\overline{D\psi}\wedge D\psi\\
&+8i\,\Sigma_{Mm}\,e^a\wedge\overline{D\psi}\wedge\gamma_5\gamma_a\psi
-8\,\Sigma_{Nm}\,e^a\wedge\overline{D\psi}\wedge\gamma_a\psi\\
&-8i\,\widetilde M\,e^a\wedge e^b\wedge\bar\psi\wedge\gamma_5\gamma_{ab}\psi
-8\,\widetilde N\,e^a\wedge e^b\wedge\bar\psi\wedge\gamma_{ab}\psi ,\\[6pt]
L^{(4)}={}&-\tfrac14\Sigma_M\,\epsilon_{abcd}\,
\bigl(\bar\psi\wedge\gamma^{ab}\psi\bigr)\wedge\bigl(\bar\psi\wedge\gamma^{cd}\psi\bigr)
-\tfrac12\Sigma_N\,
\bigl(\bar\psi\wedge\gamma^{ab}\psi\bigr)\wedge\bigl(\bar\psi\wedge\gamma_{ab}\psi\bigr)\;\equiv\;0 .
\end{align*}
\end{widetext}
The bosonic part is
\begin{equation*}
\begin{aligned}
    \mathcal{L}_{\text{bos}}=\bigl[\,384\,\Pi+\Sigma_M\,\mathcal{G}+8\,\Sigma_{M(m^2)}\,R
\\
-8\,\Sigma_{N(m^2)}\,\mathcal{H}-2\,\Sigma_N\,\mathcal{P}\,\bigr]\,\mathrm{dvol},
\end{aligned}
\end{equation*}
On the torsion constraint $R$ and $\mathcal G$ become the Riemannian scalar curvature and Gauss--Bonnet density, the Holst term $\mathcal H$ vanishes identically, and the effective cosmological constant is $\Lambda_{\rm eff}=-24\,\Pi/\Sigma_{M(m^2)}$.

The fermionic, quadratic term reads as
\begin{equation}
\begin{aligned}
\mathcal{L}_\psi ={}& 8i\,\Sigma_{Mm}\, e^a\wedge\overline{D\psi}\gamma_5\gamma_a\psi
- 8\,\Sigma_{Nm}\, e^a\wedge\overline{D\psi}\gamma_a\psi \\
&+ 4\, e^a\wedge e^b\wedge\bar\psi\gamma_{ab}\bigl(S_N - iS_M\gamma_5\bigr)\psi \\
&-8i\,d\Sigma_M\wedge\bar\psi\gamma_5D\psi+8\,d\Sigma_N\wedge\bar\psi D\psi\\
&+d\bigl(8i\Sigma_M\,\bar\psi\gamma_5D\psi-8\Sigma_N\,\bar\psi D\psi\bigr).
\end{aligned}
\label{eq:fermion}
\end{equation}
The first term is the Rarita--Schwinger action. The second one contains a torsion coupling and a coupling to $d\Sigma_{Nm}$: by
$e^a\wedge\overline{D\psi}\gamma_a\psi=\tfrac12\bigl[T^a\wedge\bar\psi\gamma_a\psi-d(e^a\wedge\bar\psi\gamma_a\psi)\bigr]$,
$-8\Sigma_{Nm}e^a\wedge\overline{D\psi}\gamma_a\psi=-4\Sigma_{Nm}T^a\wedge\bar\psi\gamma_a\psi
-4\,d\Sigma_{Nm}\wedge e^a\wedge\bar\psi\gamma_a\psi+d\bigl(4\Sigma_{Nm}e^a\wedge\bar\psi\gamma_a\psi\bigr)$. The third term is the mass term; its $N$-sector part $\propto S_N$ vanishes for one copy but not for
several copies with different masses. The fourth and fifth terms are derivative couplings of the gravitino to the
moduli, proportional to $d\phi$ and vanishing for constant $\Sigma_M$ and $\Sigma_N$. The last term is exact and does not contribute to the field equations.

Varying $L_\psi$ for $\bar\psi$ with a fixed bosonic background, one obtains $\delta\int\mathcal L_\psi=\int\delta\bar\psi\wedge\bigl(\mathcal E^{(0)}+\mathcal E^{(\partial)}\bigr)$, where
\begin{align}
&\begin{aligned}
    \mathcal{E}^{(0)} ={}& -16i\,\Sigma_{Mm}\,e^a\wedge\gamma_5\gamma_a\,\mathring{D}\psi\\
& +8\,e^a\wedge e^b \wedge\gamma_{ab}\bigl(S_N-iS_M\gamma_5\bigr)\psi ,
\end{aligned}\\[4pt]
&\begin{aligned}
\mathcal{E}^{(\partial)} ={}& 8i\,d\Sigma_{Mm}\wedge e^a\wedge\gamma_5\gamma_a\psi
-8\,d\Sigma_{Nm}\wedge e^a\wedge\gamma_a\psi\\
&+16i\,d\Sigma_M\wedge\gamma_5 D\psi
-16\,d\Sigma_N\wedge D\psi\nonumber\\
&-2i\,\Sigma_{Mm}\,e^a\wedge K^{bc}\wedge\bigl(\gamma_{bc}\gamma_5\gamma_a-\gamma_a\gamma_5\gamma_{bc}\bigr)\psi\\ 
&-8\,\Sigma_{Nm}\,T^a\wedge\gamma_a\psi .    
\end{aligned}
\end{align}
with $\mathring{D}$ being the covariant derivative of the Levi-Civita connection $\mathring\omega[e^a]$ and
$K^{ab}=\omega^{ab}-\mathring\omega^{ab}$ is the contorsion, $T^a=De^a=K^a{}_b\wedge e^b$. Define the gravitino mass $m_{3/2}$:
\begin{equation*}
m_{3/2}=\frac{\sqrt{S_M^2+S_N^2}}{|\Sigma_{Mm}|},
\end{equation*}
The supersymmetry relation $\Lambda_{\rm eff}=-3m_{3/2}^2$, i.e. $8\,\Pi\,\Sigma_{Mm}^2=\Sigma_{M(m^2)}(S_M^2+S_N^2)$, holds identically if all the $m_i$ are equal.
The construction links the moduli dynamics to the supergravity sector. A coupling to the Yang-Mills fields can be built by considering $\mathcal{N} = 2$ SUSY \cite{Alvarez_2021}.

The standard gravitino mass is calculated from the holomorphic superpotential $W(\phi,\bar\phi)$ as $m_{3/2}=e^{K(\phi ,\bar{\phi })/2}\,|W(\phi )|$. The scalar potential for spontaneously broken SUSY is given by $V(\phi,\bar\phi)=e^{K}\left(g^{I\bar{J}}D_{I}W\overline{D_{J}W}-3|W|{}^{2}\right)$. Consider an ansatz $\Lambda_{\rm eff} = V(\phi, \bar\phi)$ and $m_{3/2} = e^{K/2}|W|$ (Planck units):
\begin{equation*}
    \begin{gathered}
        -\frac{24\,\Pi}{\Sigma_{M(m^2)}} = e^{K}\left(g^{I\bar{J}}D_{I}W\overline{D_{J}W}-3|W|{}^{2}\right) \\
        \frac{\sqrt{S_M^2+S_N^2}}{|\Sigma_{Mm}|}=e^{K(\phi ,\bar{\phi })/2}\,|W(\phi )|
    \end{gathered}
\end{equation*}
Eliminating $e^{K/2}|W|$ with the K\"ahler current $D_I W/W=G_I$, the system takes the compact form:
\begin{equation*}
g^{I\bar{J}}G_{I}G_{\bar{J}}-3=-\frac{24\,\Pi\,\Sigma_{Mm}^2}{\Sigma_{M(m^2)}\,\bigl(S_M^2+S_N^2\bigr)} ,
\end{equation*}
whose right-hand side is homogeneous of degree zero in the $m^{(i)}$, so that no rescaling of the structure functions is needed. On the supersymmetric locus $8\,\Pi\,\Sigma_{Mm}^2=\Sigma_{M(m^2)}(S_M^2+S_N^2)$ it gives $G_I=0$, as unbroken supersymmetry requires.
The matrix $M$ can be chosen so the overall theory satisfies the so-called `no-scale` condition $g^{I\bar J} G_I G_J - 3 = 0$, when the scalar potential $V(\phi,\bar\phi) = 0$; this happens if and only if $\Pi=0$. The bosonic sector (on the torsion constraint, Planck units, $16\,\Sigma_{M(m^2)}=1$), will become:
\begin{equation}
\mathcal{L}_{\text{bos}}=\frac{1}{2}R - V + X +\Sigma _{M}(\phi ,\bar{\phi })\,\mathcal{G}-2\,\Sigma _{N}(\phi ,\bar{\phi })\,\mathcal{P}  
\end{equation}
where $V = m_{3/2}^2(g^{I\bar J}G_I G_{\bar J}-3) = 0$ for $m_{3/2}=e^{K/2}|W|$ being a SUSY scale (gravitino mass), and $X$ being a $\sigma$-model kinetic term. Usually, a no-go theorem is considered: for any model of $R - X + f(\phi)\mathcal{G}$ type without the potential $V(\phi)$ contains gradient instabilities \cite{Hikmawan_2016}. Two mechanisms could generate an effective potential. The first, the Holst term generating an effective K-inflation term in the presence of torsion \cite{PhysRevD.78.064070}, is not available here: on the torsion constraint the Holst density vanishes identically, and the effect belongs to the unconstrained first-order theory. The second survives: non-perturbative quantum effects for the axion term $\Sigma_N\mathcal{P}$ generate effective instantonic contributions $V(\phi) \sim \Lambda^4_G \left( 1- \cos \left(\frac\phi {f_{a}}\right)\right)$, which need no torsion. Whether they alone evade the no-go theorem remains to be shown.

\bibliography{bibliography}

\end{document}